\documentclass[12pt]{article}
\usepackage[utf8]{inputenc}
\usepackage[T1]{fontenc}
\usepackage{amsmath}
\usepackage{amssymb}
\usepackage{graphicx}
\usepackage[a4paper,width=180mm,top=20mm,bottom=20mm]{geometry}
\usepackage{cite}
\usepackage[round,sort,numbers,authoryear]{natbib}
\usepackage{authblk}
\usepackage{xcolor}

\title{Effect of Incoming Solar Particle Radiations on The Exosphere of Mars}

\author[$a\star$]{Kamsali Nagaraja}
\author[$a$]{Praveen Kumar Basuvaraj}
\author[$a$]{S.\,C. Chakravarty}
\author[$b$]{Praveen Kumar Kuttanpillai}

\affil[$a$]{Department of Physics, Bangalore University, Bengaluru, India}
\affil[$b$]{Indian Space Research Organisation Headquarters, Bengaluru, India}
\affil[$\star$]{Correspondence to : \texttt{kamsalinagaraj@bub.ernet.in}}
\date{}

\begin{document}

\maketitle

\begin{abstract}
Mars Exospheric Neutral Composition Analyzer (MENCA) of Mars Orbiter Mission (MOM) measures the \emph{in-situ} neutral upper atmospheric constituents of Mars. Martian lower atmosphere predominated by the presence of $CO_2$ which photo-dissociates into atomic oxygen ($O$) in higher altitudes much near the exobase. Atomic $O$ plays a significant role in invoking stronger presence of $O_2^+$ in the Martian ionosphere. Primary photo-dissociative species $CO_{2}$, crossover its neutral abundance with atomic $O$ in the collisonless hetergenous atmosphere with varying local solar conditions. Initial measurements from Neutral Gas and Ion Mass Spectrometer (NGIMS) instrument on Mars Atmosphere and Volatile Evolution (MAVEN) estimated these crossover/transition altitude wavering between $\approx$225\,km to 240\,km during solar maximum conditions with peak solar illuminations. MENCA sampled the neutral atmospheric species, below the exobase upto periareion of $\approx$160\,km, under low solar active conditions during June 2018. Observations of partial pressures of $CO_2$ and $O$ in subsequent orbits reveals that solar inputs are crucial in quantifying these crossing points, where $[O]/[CO_2]$ remain unity, alongside the influences from temperature. The multi-spacecraft measurements of the direct influences of solar wind charged particle fluxes and velocity on the daily variation of neutral thermospheric/exospheric compositions were observed on the local evening hours of Mars and presented. It marks the first-ever direct \emph{in-situ} observation of interaction between the energetic solar particle radiations on Martian exospheric compositions, potentially contibuting for the steady escape and differing population of atomic $[O]$ in the exosphere.
\end{abstract}


\section{Introduction}
The seasonal cycles of summer, autumn, winter and spring on Mars are similar to that of Earth due to their near similar axial tilts (25$^{\circ}$ for Mars and 23.5$^{\circ}$ for Earth), but each season lasts for almost double the time on Mars as compared to Earth because of the difference in their periods of revolution around the Sun. Along with seasons, parameters such as solar extreme ultraviolet fluxes, radiative and collisonal cooling, gravity waves, heliocentric distance, laitute, etc., would have influenced the upper atmospheric processes on Mars \citep{Bougher2015AeronomyOfMars_ACC-MAVEN}. For several decades, the results on upper atmospheric composition and density of Mars were limited to the entry, descent and landing (EDL) observations carried out by the Viking Lander 1/2 that reached Martian surface on 20 July 1976 at 22.5$^{\circ}$N, 48$^{\circ}$W and 03 September 1976 at 48$^{\circ}$N, 22$^{\circ}$W respectively. These \emph{in-situ} measurement of atmospheric profile below 200 km confirmed the presence of abundant $CO_2$ supporting earlier Ultra-Violet Spectroscopic (UVS) observation using Mariner 6, 7 and 9 missions \citep{Barth1971_MUA_UVS-Mariner6-7,Barth1972UVS_AirGlowLyman_Variation_UVS-Mariner9,Stewart_Barth1972StrucOfMUA_UVS-Mariner9}. Alongside $CO_2$, the presence of $N_2$, $Ar$, and trace measurements of atomic and molecular oxygen ($O$ and $O_2$), $CO$, $NO$, $Kr$, $Ne$ and $Xe$ were reported \citep{Nier1976StrucOfMNUA_NMS_Viking1-2,Nier1977CompStrucOfMUA_NMS-Viking1-2,Owen+Biemann1976CompOfMarsAtmosSurface_Argon36Detection,Owen1977CompOfMarsAtmosSurface}. The complete set of near surface meteorological data obtained from Viking Landers (1976) to the Curiosity Rover (2012) has been analysed and results have been consolidated in terms of diurnal, seasonal and inter-annual variations of meteorological parameters including dust storms over a span of more than 20~Martian years \citep{Martinez2017NearSurfaceMarsAtmos_VikingToCuriosity}. Though, the Viking missions succesfully identified the major and minor species of Martian atmosphere to surface, it is to be noted that this observational data provides paucity on the study pertaining to seasonal and solar activity related variations of the Martian thermosphere and exosphere. A similar analysis to characterise the thermosphere/exosphere system has not been possible due to the lack of observational data on the atmospheric neutral/ionised gas constituents covering the exospheric region \citep{Bougher2015AeronomyOfMars_ACC-MAVEN}. 

Indian Space Research Organisation (ISRO) has launched its first interplanetary mission to Mars \emph{viz.} Mars Orbiter Mission (MOM) on 05 November 2013, reached Mars on 24 September 2014 during Martian nighttime. MOM arrived at Mars with its initial primary objective concerning studies on planet's morphological features, detection of Methane ($CH_4$), estimating the ratio between Deuterium and atomic Hydrogen ($D/H$), coverage of spatio-temporal profiles of various neutral atmospheric constituents and cartographic events including dust storms, atmospheric clouds and so-on \citep{KiranKumar2014ScientificExplorationOfMOM}. As of August 2020, the MOM spacecraft remains healthy in performing the extended mission objectives.

Utilizing the opportunity that Mars approaches Earth closely, for every $\approx$2 Gregorian years, in collateral to the Mars Orbiter Mission, National Aeronautics and Space Administration (NASA) launched the Mars Atmosphere and Volatile Evolution (MAVEN) mission on 18 November 2013 reaching Mars on 22 September 2014. MAVEN mission dedictated to study the structure, composition and dynamics of the upper atmosphere of Mars (above $\approx$150\,km), also investigated the role of solar wind plasma and habitability aspects of Mars \citep{Jakosky2015MAVEN}. The ExoMars Trace Gas Orbiter (ExoMars TGO) mission launched on 14 March 2016, as an joint venture between the European Space Agency (ESA) and Roscosmos (Russian Space Agency), inserted intially into a periareion of $\approx$400\,km exospheric orbit at an inclination of 74$\circ$, much similar to MAVEN's orbit inclination, for observing biologically relevant trace constituents such as Methane ($CH_4$) and potential organic gases \citep{Olsen2017ACE-MIRsuite_ExoMarsTGO}.

Based on the observations using MENCA payload between December 2014 and May 2015, analysis to extract spatial and temporal distribution of the atmospheric composition of the thermosphere-exosphere region of Mars have been carried out by \citet{Nagaraja2020ExosphereOfMars_MENCA-MOM}. The results obtained from NGIMS-MAVEN data have provided valuable information about the spatial variation of the thermospheric neutral/ion constituents delineating their vertical and latitudinal distribution, and the effect of solar zenith angle \citep{Mahaffy2015StrucCompMNUA_NGIMS-MAVEN}. MAVEN Deep Dip campaign to sample the sub-solar collisional homosphere and magnetic field structure in April 2015 demonstrated the orbit-to-orbit variation of thermosphere and ionosphere of Mars \citep{Bougher2015EarlyDeepDipCampaign_MAVEN}. Direct measurements of atmospheric neutrals from MENCA-MOM and NGIM-MAVEN has been used in our study to examine the influence of solar activity driven changes on Martian upper atmosphere. The results from both the spacecrafts agrees that the energetic particle fluxes deposited over the exosphere plays a major role in steering the daily variations on the exospheric compositions of Mars.

\section{Data Analysis and Methodology}
\subsection{Mars Orbiter Mission}

Mars Orbiter Mission spacecraft inserted into Martian orbit on 24 September 2014, initially with a periareion of $\approx$400\,km. MOM had a orbital period of $\approx$72~hours due to its highly eccentric orbit inclined at 150$^{\circ}$ focusing observations over the Martian equator. Comet Siding Spring (C/2013~A1) had a closest approach by Mars on 19~October~2014, within a month of arrival of MOM and MAVEN spacecrafts at Mars. Operations of all spacecrafts were ceased to protective mode to prevent from the meteor shower caused during the passage of the comet \citep{Schneider2015AftermathOfCometSidingSpring_IUVS-MAVEN}. MOM occulted behind Mars through orbital manoeuvres, later brought down its periareion altitudes to around 262\,km during December~2014, more feasible region to study the Martian exosphere.

\begin{figure}[h]
\begin{center}
		\noindent\includegraphics[width=0.75\textwidth]{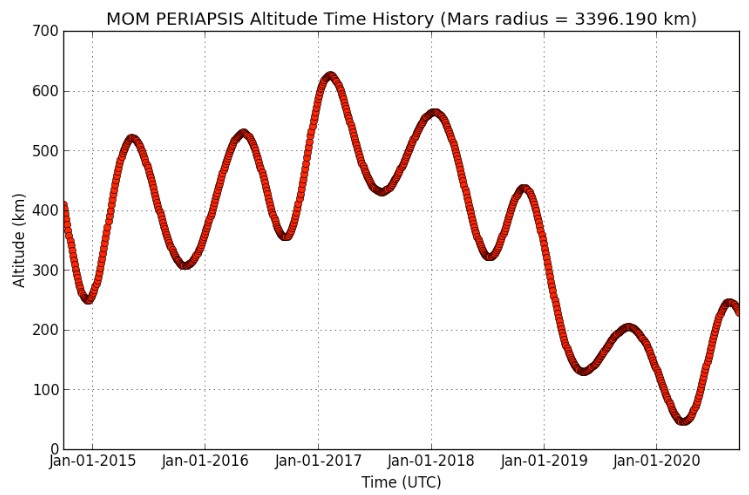}
\end{center}
	\caption{Mars Orbiter Mission's periareion altitudes between October~2014 and October~2020, as projected by the MOM team at Jet Propulsion Laboratory. The figure forecasts the periareion altitudes reached by MOM immediate after orbital maneuvers due to the closest encounter of Comet Siding Spring (C/2013~A1) by Mars on 19~October~2014 \citep{Helfrich2015JourneyWithMOM}.}
	\label{fig01}
\end{figure}

Mars Exospheric Neutral Composition Analyser (MENCA), a dedicated atmospheric suite of MOM, sampled exospheric neutrals near the exobase from December~2014 through May~2015 comprising of 88~orbits. MENCA measures the total atmospheric pressure and partial pressures of various atmospheric constituents covering the mass range from 1-300~\emph{amu}, which is further programmable to limit its operation under mass-sweep and trend modes. MENCA consists of a built-in electron impact ionizer operated at $\approx$70~eV to ionize the atmospheric neutrals, a set of four quadrupole rods and a detector assembly. The detector comprises of Faraday Cup and Channel Electron Multiplier measures the partial pressures of gas constituents with a mass resolution of 1 \emph{amu}. Part of postively ionized atmospheric neutrals inside the ionization chamber known as the source grid region drags the ions inside the quadrupole mass filter system. Another part of neutrals ionized outside the source grid were collected by the Bayard-Alpert gauge, which is calibrated to measure the total pressure. The detailed description of the instrumentation, calibration factors, complete working mechanism, sensitivity and measurement limitations were given by \citet{Bhardwaj2016EveningExosphereOfMars_MENCA-MOM}.  

Indian Space Science Data Centre (ISSDC) at Bengaluru, India archives the MOM observational data and disseminate to scientific users. This archived data consists of total pressure and partial pressure values in the units of Torr with variable time resolution of $\approx$12 to 30\,s. The data is used for scientific studies after incorporating calibration, correction and normalization factors, and time tagging the processed data with ancillary information such as latitude, longitude, altitude and solar zenith angle (SZA) corresponding to the orbital phase of MENCA-MOM using relevant SPICE/Ephemeris kernels \citep{Nagaraja2020ExosphereOfMars_MENCA-MOM}. The data sets are identified and arranged with respect to different orbit numbers of MOM. However due to the oscillatory nature of periareion height of MOM, good coverage of lower and crucial exospheric altitudes has been obtained only for a small number of orbits (in spite of some orbit lowering excercises) during 5-13,~June~2018.

Figure \ref{fig01} shows the periareion altitude forecast of MOM spacecraft from October~2014 through October~2020. It is clear that after 2014-2015 there has been only a very short period during 2018-2019 when useful data could be collected closer to the exobase and above. The predicted minimum periareion altitude reached by MOM will be 112.9\,km on 08~July~2029 and can traverse above 100\,km \citep{Helfrich2015JourneyWithMOM}. The analysis was carried out for the additional MENCA data available during June~2018 and results obtaiend are discussed in detail.

\subsection{Mars Atmosphere and Volatile Evolution}
The Mars Atmosphere and Volatile EvolutioN (MAVEN) mission inserted into a highly elliptical orbit, focuses on the search for past history of Martian atmosphere and to understand its present climate. The Neutral Gas and Ion Mass Spectrometer (NGIMS) instrument of the MAVEN spacecraft studies the structure and composition of the upper neutral atmosphere of planet Mars, measures isotope ratios, and measures thermal and supra-thermal ions \citep{Mahaffy2015NGIMS-MAVEN}. NGIMS measures in a mass range of 2-150 \emph{amu}. NGIMS science operation starts below 500\,km to periapsis (inbound) and periapsis to 500\,km (outbound) during each orbit lasting for $\approx$600 s for each leg, with spatial resolution of 1\,km. These observation covers both exosphere as well as thermosphere of Martian atmosphere that peers through the exobase at $\approx$200\,km, falls within NGIMS limits. The level-2 (version 08 and revision 01) data-sets of NGIMS-MAVEN has been retrieved from MAVEN Science Data Center at Laboratory for Atmosphere and Space Physics (LASP), during the period June~2018 were used for this study. NGIMS-MAVEN observation has been been chosen from 03~June~2018 through 15~June~2018 in parallel to MENCA-MOM observations.

\section{Results and Discussion}

\begin{figure}[h!]
\begin{center}
	\includegraphics[width=0.75\textwidth]{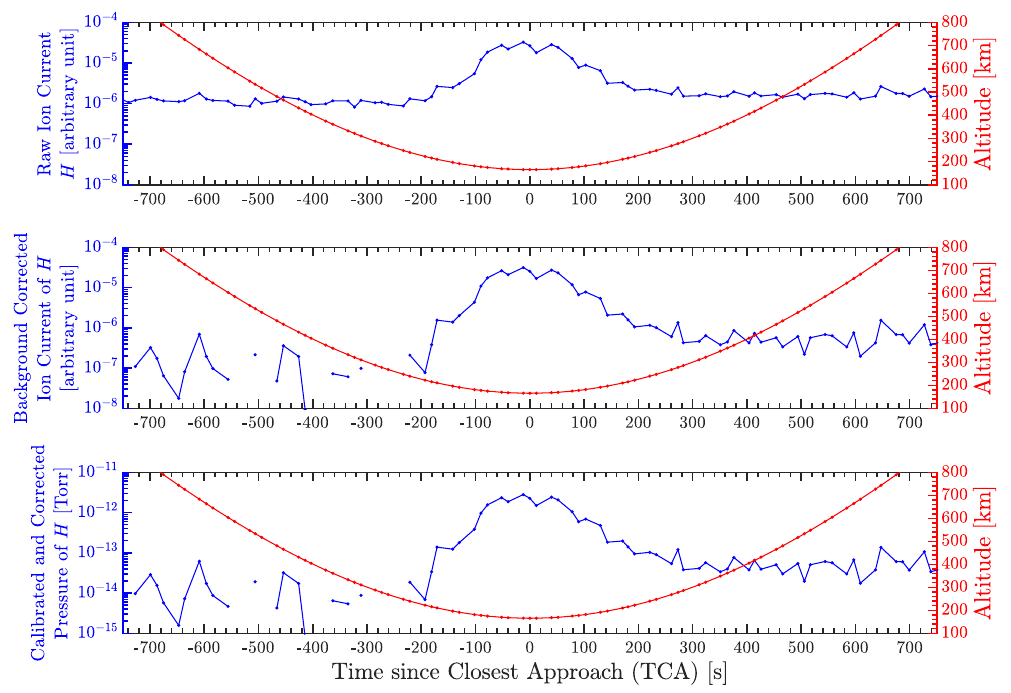} \end{center}
	\caption{Spatio-temporal variation of raw ion current and calibrated partial pressure of atomic Hydrogen ($H$) observed from MENCA-MOM on 05~June~2018.}
	\label{fig02}
\end{figure}

Figure \ref{fig02} shows the analysis of atomic $H$ ion current data on 5~June~2018 covering the time period 10 to 13\,h UTC having both temporal and spatial variations. The time period of closest approach \emph{i.e.} near periareion, of MOM towards Mars has been shown along with altitude coverage. The top panel shows altitude profile derived from SPICE kernels provided along with the observational data and ion current of raw data. The middle panel shows the ion current of atomic $H$ with background corrected at $\approx$500\,km. The bottom panel shows the calibrated partial pressure of atomic $H$ for the actual time sampled values with background correction. The pressure values have been converted to Torr units by using the calibration procedure prescribed in \citet{Bhardwaj2016EveningExosphereOfMars_MENCA-MOM}. The termainal panel plot clearly shows that the maximum pressure values occur much near to the periareion altitudes though it can vary from orbit-to-orbit. The figure also shows a small height interval near pariareion with increasing pressure much above the noise levels of the instrument. This is the region of useful data to be used for further analysis.

From the ion currents differentiated with \emph{amu} values, the parial pressures of atmospheric constituents can be estimated for obtaining the time/altitude profiles. Since the traverse of spacecraft through the periareion is relaively of shorter time to collect useful data within same orbit, the effect of change in solar zenith angle may be insignificant. However, for long-term variabilty, the seasonal changes in solar zenith angles need to be corrected for studying any effect of solar activity related variations in exospheric partial pressures.

\begin{figure}[h!]
	\begin{center}
		\includegraphics[width=0.75\textwidth]{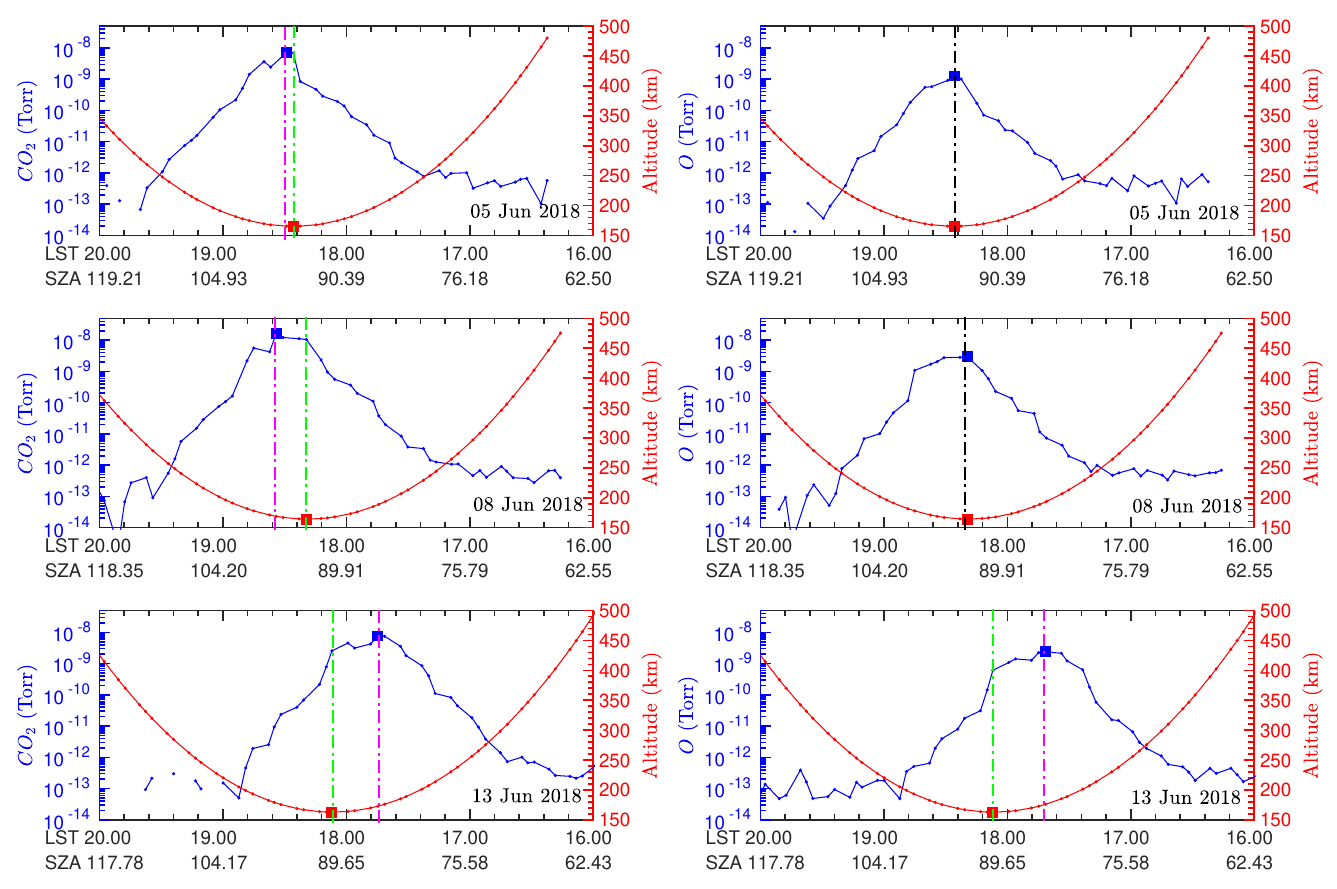}
	\end{center}
	\caption{Variation of atmospheric $CO_2$ and $O$ abundances observed on 05, 08 and 13~June~2018 by MENCA-MOM. Corresponding variation in local solar time (LST) in hours and solar zenith angle (SZA) in degrees has been shown.}
	\label{fig03}
\end{figure}

Figure \ref{fig03} shows details of the variation of atmospheric $CO_2$ and $O$ partial pressures near the periareion region between $\approx$160\,km to 500\,km, subtracting the background noise at 500\,km. This covers the Martian upper thermosphere, its exobase and lower exosphere regions for ascertaining relative effects of gaseous transition and escape. While the pattern of variation of partial pressure is similar on 05~June 2018 and 08~June~2018, there is a shift of about few minutes in the peak density observed on 13~June~2018, where maximum of pressure has not occurred at the lowest height of observation in both $CO_2$ and atomic $O$. This anomaly needs to be examined by studying the condition of atmospheric dynamics. It is also observed that the rate of change of pressure values for inbound and outbound trajectories are generally symmetric. It can be seen that the SZA variation is from the nighttime and moving towards the evening hours of Mars. This transition does not affect the density profiles appreciably for short duration of observations.

\begin{figure}[h!]
	\begin{center}
		\includegraphics[width=0.75\textwidth]{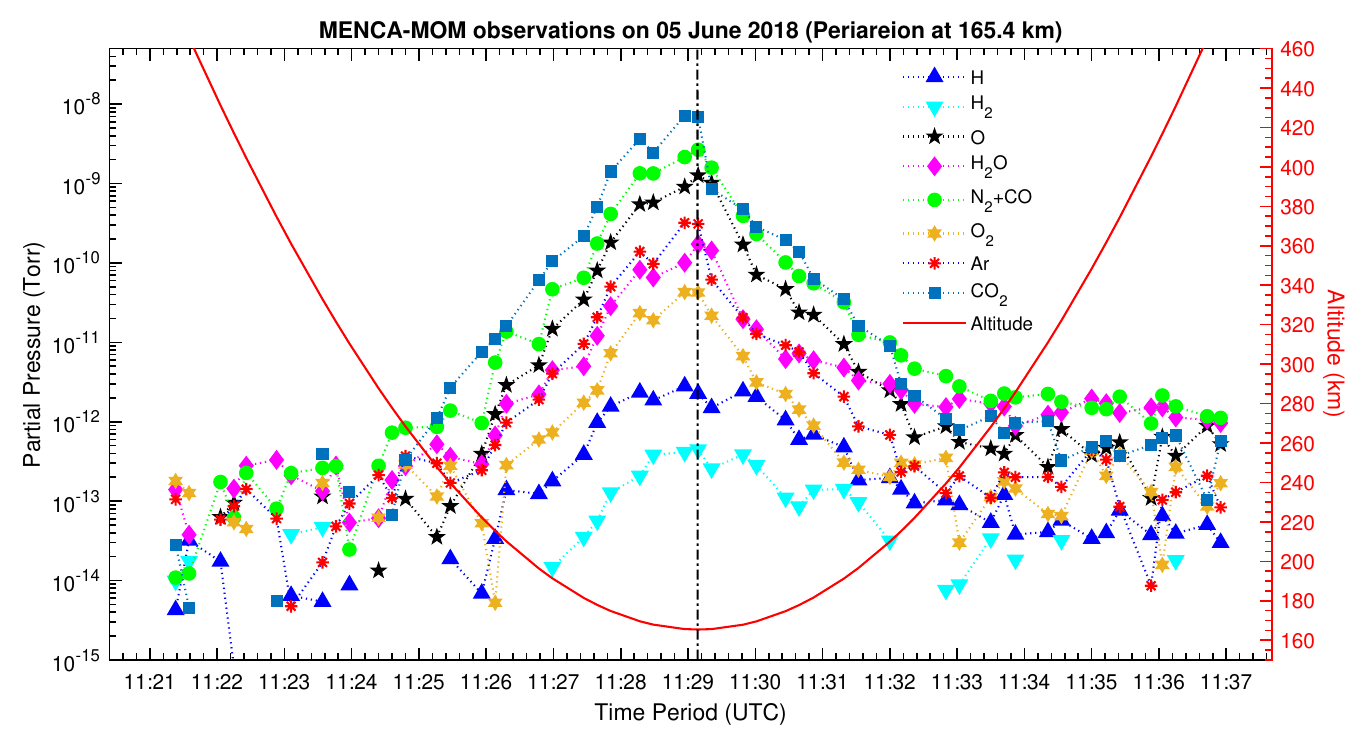}
	\end{center}
	\caption{Relative abundance of major atmospheric species measured from MENCA-MOM on 05~June~2018. Increased $N_2+CO/amu28$ has contributions from $CO_2/amu44$. Variation of water vapour ($H_2O$) is mainly due to the outgassing of MOM spacecraft and eliminated during background correction.}
	\label{fig04}
\end{figure}

\begin{figure}[h!]
\begin{center}
		\includegraphics[width=0.75\textwidth]{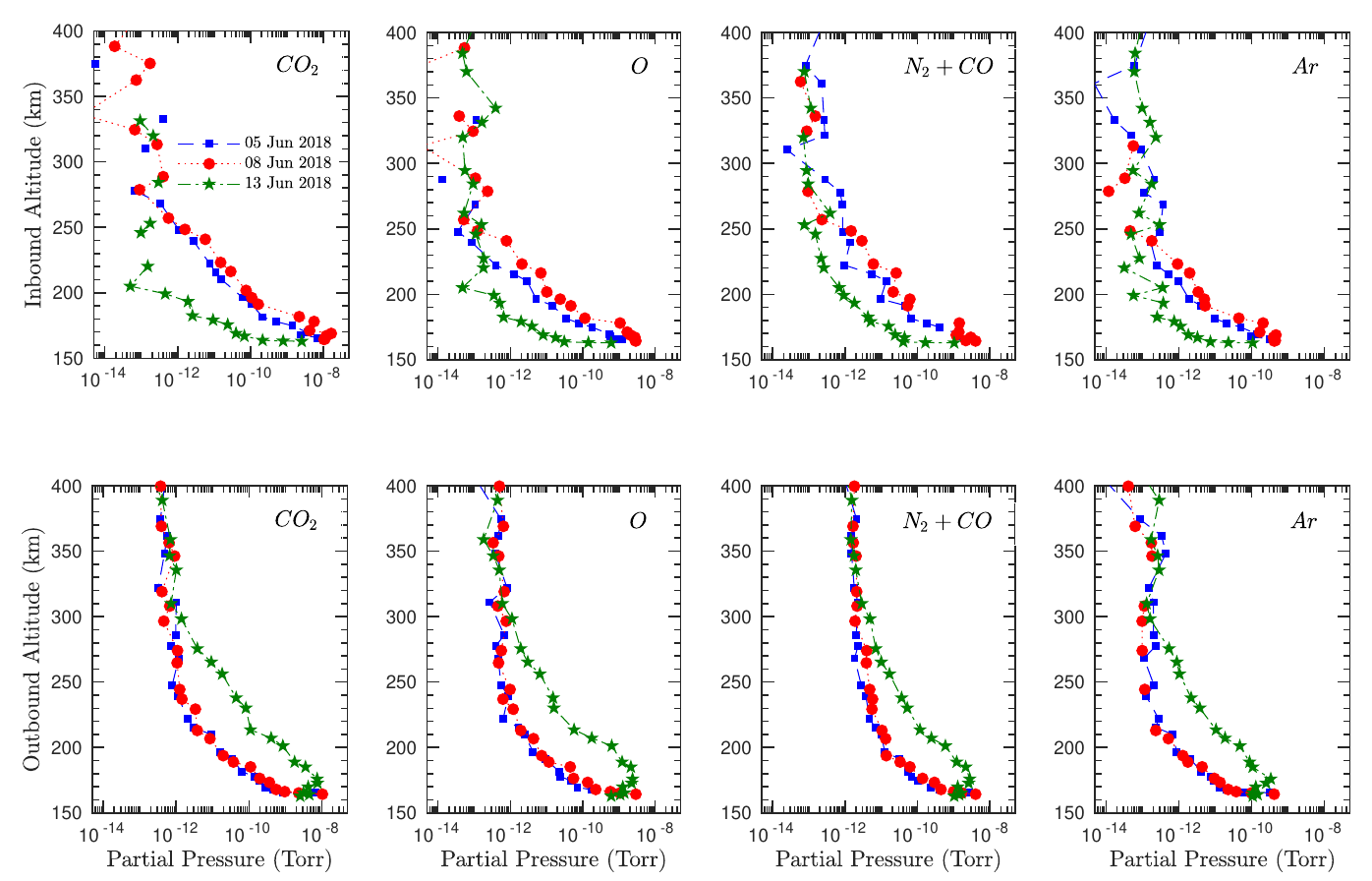}
\end{center}
	\caption{Inbound (top) and Outbound (bottom) altitude profiles of $CO_2$, $O$, $CO+N_2$ and $Ar$ on 05, 08 and 13~June~2018 from MENCA of Mars Orbiter Mission.}
	\label{fig05}
\end{figure}

\begin{figure}[h!]
\begin{center}
		\includegraphics[width=0.75\textwidth]{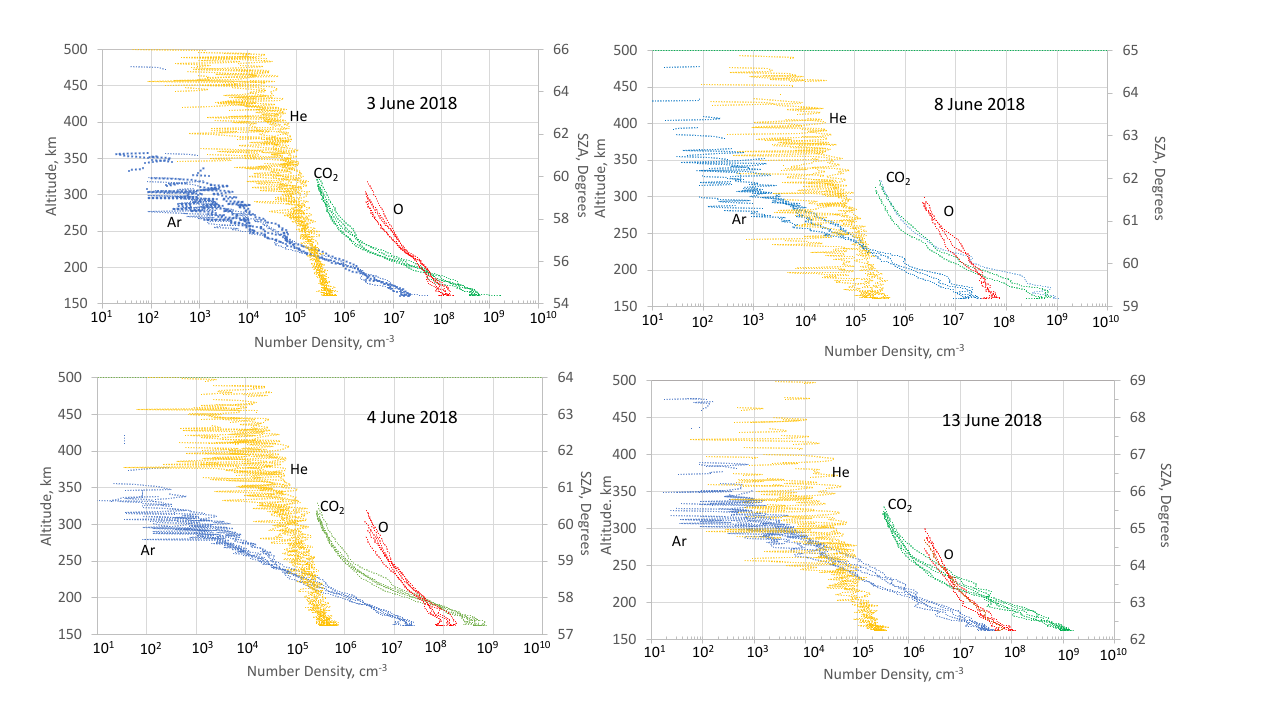}
\end{center}
	\caption{Vertical profiles of upper atmospheric neutral gas constituents for four days \emph{viz.,} 03, 04, 08 and 13~June~2018. The corresponding variations of SZA during the period of each profiles are also shown.}
	\label{fig06}
\end{figure}

Figure \ref{fig04} shows the relative partial pressure of major atmospheric constituents. It is noted that the $CO_2$ shows the maximum density values. According to the known pattern of variations, the other constituents have relatively lower pressures compared to $CO_2$. In addition, the water vapor density, product of outgassing, in MOM has reduced considerably compared to the values observed during December~2014. However, the temporal profile of $H_2O$ follows similar pattern of other species due to reduction in further degassing. The escaping remanant water vapour may be maintained through the hydrostatic equilibrium.

To undersand the day-to-day variations of the partial pressures of the exospheric constituents, the profiles of $CO_2$, $Ar$, $O$ and $N_2+CO$ are plotted for the three orbits of observations in Figure \ref{fig05}. It is seen that during a short period of nine days, there is considerable reduction in pressure on 13~June~2018 compared to 05 June 2018 and 08 June 2018. Such short period changes of gas concentrations in the thermosphere/exosphere need to be first confirmed before extending any possible explanation. To strengthen MENCA observations and analysis, the simultaneous observations from MAVEN on Martian thermosphere/exosphere were used. NGIMS-MAVEN atmospheric payload measures the neutral/ions of upper atmosphere has been utilised to validate the observed variations. A detailed analysis was carried out using NGIMS-MAVEN data during 03-14,~June~2018 to match period of MENCA-MOM. There are about five orbits of MAVEN per Earth day with NGIMS data covering the Martian atmospheric altitude between $\approx$150\,km and 500\,km taking about 11-min for the track/traverse. The daily average values of parameters are computed from the available data during the 24\,h of each Earth day. Figure \ref{fig06} shows the results of the altitude profiles of a few selected Martian atmospheric constituents in the units of number density per cubic centimeter for four days of June~2018.

\begin{figure}[h!]
\begin{center}
		\includegraphics[width=0.75\textwidth]{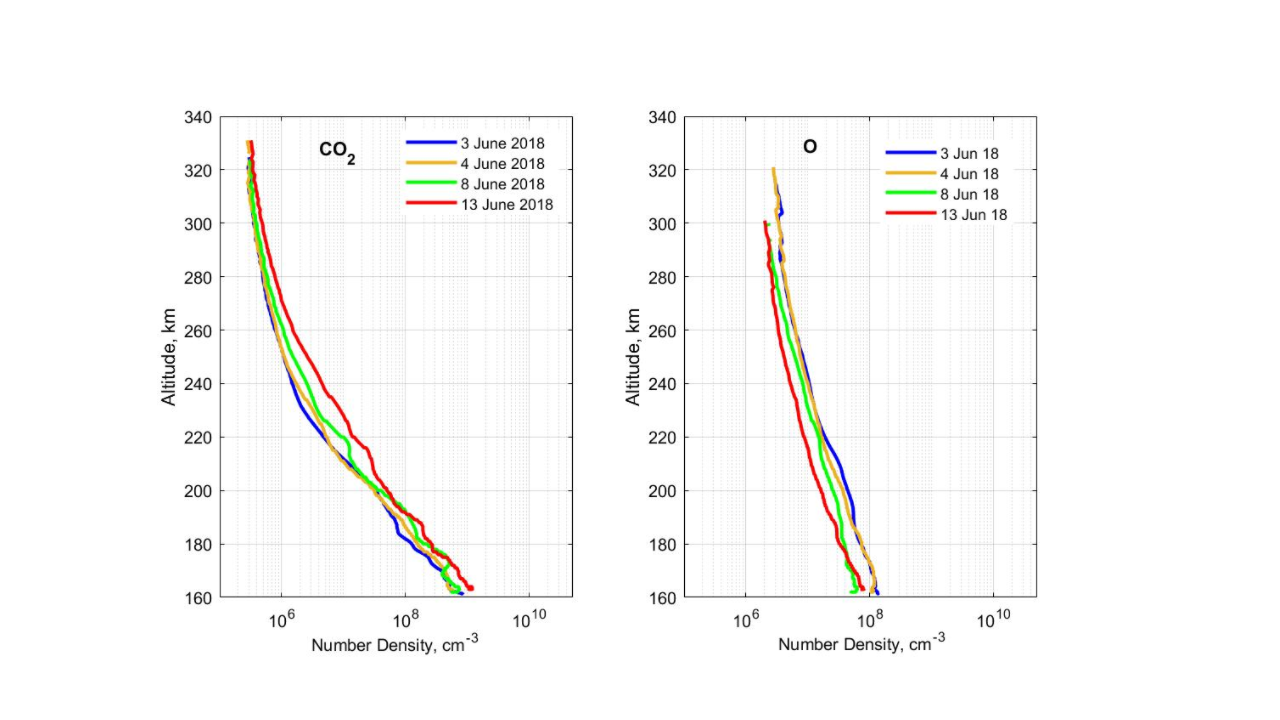}
\end{center}
	\caption{Vertical profiles of daily mean conentrations of $CO_2$ and $O$ densities on 03, 04, 08 and 13~June~2018. [$CO_2$] steadily increase with decline in [$O$] during this period as observed from NGIMS-MAVEN.}
	\label{fig07}
\end{figure}

\begin{figure}[h!]
	\begin{center}
		\includegraphics[width=0.60\textwidth,angle=-90]{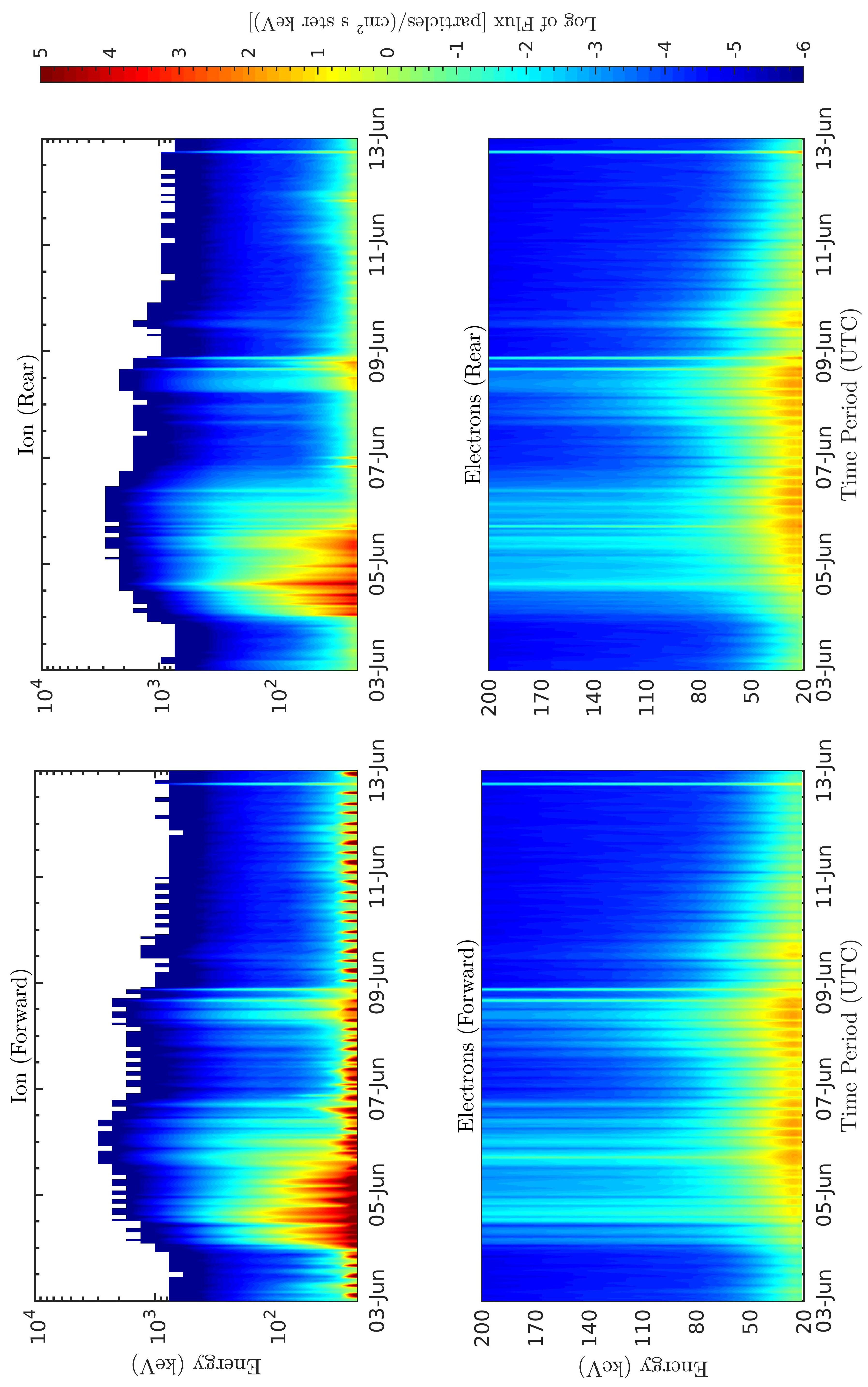}
	\end{center}
	\caption{Ion and electron fluxes signifying the solar-wind variations as measured from the SEP instrument (forward and rear-view) of MAVEN during 03-13, June 2018. Steady reduction in the incoming solar flux, immediate after 05~June~2018, shows significant contribution to the decrease of the photodissociative product, atomic oxygen ($O$).}
	\label{fig08}
\end{figure}

On 13~June~2018, differences in the density profiles can be seen compared to the profiles on 03, 04 and 08~June~2018 that are similar. Particularly, the [$CO_2$] and [$O$] crossover point has shifted to a higher altitude of $\approx$240\,km on 13~June~2018 compared to $\approx$190\,km on other days. This indicates a reduction in photodissociation driven concentration of atomic oxygen ($O$) on 13~June~2018. Such reduction of atomic oxygen concentrations has been observed by MENCA also (see Figure \ref{fig05}). The independent results from two separate spacecrafts confirms the anomalies which could have resulted from the solar wind particle velocity and flux variations. Figure \ref{fig07} shows the mean daily profiles of $CO_2$ and $O$. There exists a striking relationship between the decrease in $O$ and increase in $CO_2$ densities during the period from 03~June~2018 to 13~June~2018.

The Solar Energetic Particle (SEP) instrument onboard MAVEN measured the solar-wind electron and ion fluxes over various energy levels during the same period. The SEP sensors are positioned on two corners to ensure that the field of view (FOV) adequately cover the canonical Parker spiral direction around which solar energetic particle distributions are normally centered. SEP provides measurements with $\approx$1 h time resolution. The data of the electron and ion fluxes have been analysed from 03-13,~June~2018 and are shown in Figure \ref{fig08}. The energy fluxes of both ion and electron were reduced from 09~June~2018. This reduction in solar heating of the upper atmosphere resulted in the fall of photoionization of $CO_2$.
Further, the dissociation of $CO_2$ under this condition has limited the prodution of $O$, and the same is reflected.

\section{Summary}
The upper atmosphere of Mars has been observed from MENCA-MOM and NGIMS-MAVEN instruments during June 2018.
Interesting, the solar wind particle fluxes show a steady decrease between 09 June 2018 and 13 June 2018, providing the possible clues of charged particle interactions with Martian atmospheric constituents playing a major role in the day-to-day variation of $O$ and $CO_2$ concentrations. This study shows the increase in concentration of $O$ and decrease in $CO_2$ during 03-08,~June~2018 indicates an enhanced production of $O$ from the dissociation of $CO_2$ corresponding to higher energy and fluxes of solar wind particles. Hence, it marks the first-ever observation that in absence of a global magnetic field of Mars, the direct interaction of solar wind charged particles affect the daily variation of thermosphere/exosphere gaseous concentrations, contributing to the steady escape of $O$ with enhanced solar activity.

\section*{Acknowledgement}
This work was funded by the Indian Space Research Organisation (ISRO) under Mars Orbiter Mission's Announcement of Opportunity Program through the research project, Observation and Modeling Studies of the Atmospheric Composition of Mars (OMAC), vide reference number ISRO:SPL:01.01.33/16. We greatly acknowledge the use of MENCA data from MOM, archived at Payload Operation Center, Space Physics Laboratory, Vikram Sarabhai Space Centre, Thiruvananthapuram, India. The NGIMS and SEP datasets of MAVEN used for this study were publicly available on MAVEN Science Data Center at LASP (https://lasp.colorado.edu/maven/sdc/public/) as well as the Planetary Data System (http://pds.nasa.gov). The MAVEN mission is supported by NASA through the Mars Exploration Program. We thank AMDA Science Analysis System (http://amda.irap.omp.eu/) provided by the Centre de Données de la Physique des Plasmas (CDPP) supported by CNRS, CNES, Observatoire de Paris and Université Paul Sabatier, Toulouse for performing initial MAVEN data analysis on NGIMS and SEP instruments. This research work was carried at Atmospheric and Space Science Research Lab, Department of Physics, Bangalore University, Bengaluru, India.

\bibliography{references}

\begin{thebibliography}{}

\bibitem[Barth et~al., 1971]{Barth1971_MUA_UVS-Mariner6-7}
Barth, C., Hord, C., Pearce, J., Kelly, K., Anderson, G., and Stewart, A.
  (1971).
\newblock Mariner 6 and 7 ultraviolet spectrometer experiment: Upper atmosphere
  data.
\newblock {\em Journal of Geophysical Research}, 76(10):2213--2227.

\bibitem[Barth et~al., 1972]{Barth1972UVS_AirGlowLyman_Variation_UVS-Mariner9}
Barth, C., Stewart, A., Hord, C., and Lane, A. (1972).
\newblock Mariner 9 ultraviolet spectrometer experiment: Mars airglow
  spectroscopy and variations in lyman alpha.
\newblock {\em Icarus}, 17(2):457 -- 468.

\bibitem[Bhardwaj et~al., 2016]{Bhardwaj2016EveningExosphereOfMars_MENCA-MOM}
Bhardwaj, A., Thampi, S.~V., Das, T.~P., Dhanya, M.~B., Naik, N., Vajja, D.~P.,
  Pradeepkumar, P., Sreelatha, P., Supriya, G., K., A.~J., Mohankumar, S.~V.,
  Thampi, R.~S., Yadav, V.~K., Sundar, B., Nandi, A., Padmanabhan, G.~P., and
  Aliyas, A.~V. (2016).
\newblock On the evening time exosphere of mars: Result from menca aboard mars
  orbiter mission.
\newblock {\em Geophysical Research Letters}, 43(5):1862--1867.

\bibitem[Bougher et~al., 2015a]{Bougher2015AeronomyOfMars_ACC-MAVEN}
Bougher, S., Cravens, T., Grebowsky, J., and Luhmann, J. (2015a).
\newblock The aeronomy of mars: Characterization by maven of the upper
  atmosphere reservoir that regulates volatile escape.
\newblock {\em Space Science Reviews}, 195(1-4):423--456.

\bibitem[Bougher et~al., 2015b]{Bougher2015EarlyDeepDipCampaign_MAVEN}
Bougher, S., Jakosky, B., Halekas, J., Grebowsky, J., Luhmann, J., Mahaffy, P.,
  Connerney, J., Eparvier, F., Ergun, R., Larson, D., McFadden, J., Mitchell,
  D., Schneider, N., Zurek, R., Mazelle, C., Andersson, L., Andrews, D., Baird,
  D., Baker, D.~N., Bell, J.~M., Benna, M., Brain, D., Chaffin, M., Chamberlin,
  P., Chaufray, J.-Y., Clarke, J., Collinson, G., Combi, M., Crary, F.,
  Cravens, T., Crismani, M., Curry, S., Curtis, D., Deighan, J., Delory, G.,
  Dewey, R., DiBraccio, G., Dong, C., Dong, Y., Dunn, P., Elrod, M., England,
  S., Eriksson, A., Espley, J., Evans, S., Fang, X., Fillingim, M., Fortier,
  K., Fowler, C.~M., Fox, J., Gr{\"o}ller, H., Guzewich, S., Hara, T., Harada,
  Y., Holsclaw, G., Jain, S.~K., Jolitz, R., Leblanc, F., Lee, C.~O., Lee, Y.,
  Lefevre, F., Lillis, R., Livi, R., Lo, D., Ma, Y., Mayyasi, M., McClintock,
  W., McEnulty, T., Modolo, R., Montmessin, F., Morooka, M., Nagy, A., Olsen,
  K., Peterson, W., Rahmati, A., Ruhunusiri, S., Russell, C.~T., Sakai, S.,
  Sauvaud, J.-A., Seki, K., Steckiewicz, M., Stevens, M., Stewart, A. I.~F.,
  Stiepen, A., Stone, S., Tenishev, V., Thiemann, E., Tolson, R., Toublanc, D.,
  Vogt, M., Weber, T., Withers, P., Woods, T., and Yelle, R. (2015b).
\newblock Early maven deep dip campaign reveals thermosphere and ionosphere
  variability.
\newblock {\em Science}, 350(6261).

\bibitem[Helfrich et~al., 2015]{Helfrich2015JourneyWithMOM}
Helfrich, C., Berry, D.~S., Bhat, R., Border, J., Graat, E., Halsell, A.,
  Kruizinga, G., Lau, E., Mottinger, N., Rush, B., et~al. (2015).
\newblock A journey with mom.
\newblock International Symposium on Space Flight Dynamics.

\bibitem[{Jakosky} et~al., 2015]{Jakosky2015MAVEN}
{Jakosky}, B.~M., {Lin}, R.~P., {Grebowsky}, J.~M., {Luhmann}, J.~G.,
  {Mitchell}, D.~F., {Beutelschies}, G., {Priser}, T., {Acuna}, M.,
  {Andersson}, L., {Baird}, D., {Baker}, D., {Bartlett}, R., {Benna}, M.,
  {Bougher}, S., {Brain}, D., {Carson}, D., {Cauffman}, S., {Chamberlin}, P.,
  {Chaufray}, J.~Y., {Cheatom}, O., {Clarke}, J., {Connerney}, J., {Cravens},
  T., {Curtis}, D., {Delory}, G., {Demcak}, S., {DeWolfe}, A., {Eparvier}, F.,
  {Ergun}, R., {Eriksson}, A., {Espley}, J., {Fang}, X., {Folta}, D., {Fox},
  J., {Gomez-Rosa}, C., {Habenicht}, S., {Halekas}, J., {Holsclaw}, G.,
  {Houghton}, M., {Howard}, R., {Jarosz}, M., {Jedrich}, N., {Johnson}, M.,
  {Kasprzak}, W., {Kelley}, M., {King}, T., {Lankton}, M., {Larson}, D.,
  {Leblanc}, F., {Lefevre}, F., {Lillis}, R., {Mahaffy}, P., {Mazelle}, C.,
  {McClintock}, W., {McFadden}, J., {Mitchell}, D.~L., {Montmessin}, F.,
  {Morrissey}, J., {Peterson}, W., {Possel}, W., {Sauvaud}, J.~A., {Schneider},
  N., {Sidney}, W., {Sparacino}, S., {Stewart}, A.~I.~F., {Tolson}, R.,
  {Toublanc}, D., {Waters}, C., {Woods}, T., {Yelle}, R., and {Zurek}, R.
  (2015).
\newblock The mars atmosphere and volatile evolution (maven) mission.
\newblock {\em Space Science Reviews}, 195(1-4):3--48.

\bibitem[Kumar and Chauhan, 2014]{KiranKumar2014ScientificExplorationOfMOM}
Kumar, A. S.~K. and Chauhan, P. (2014).
\newblock Scientific exploration of mars by first indian interplanetary space
  probe: Mars orbiter mission.
\newblock {\em Current Science}, 107:1096--1097.

\bibitem[Mahaffy et~al., 2015a]{Mahaffy2015StrucCompMNUA_NGIMS-MAVEN}
Mahaffy, P.~R., Benna, M., Elrod, M., Yelle, R.~V., Bougher, S.~W., Stone,
  S.~W., and Jakosky, B.~M. (2015a).
\newblock Structure and composition of the neutral upper atmosphere of mars
  from the maven ngims investigation.
\newblock {\em Geophysical Research Letters}, 42(21):8951--8957.

\bibitem[Mahaffy et~al., 2015b]{Mahaffy2015NGIMS-MAVEN}
Mahaffy, P.~R., Benna, M., King, T., Harpold, D.~N., Arvey, R., Barciniak, M.,
  Bendt, M., Carrigan, D., Errigo, T., Holmes, V., et~al. (2015b).
\newblock The neutral gas and ion mass spectrometer on the mars atmosphere and
  volatile evolution mission.
\newblock {\em Space Science Reviews}, 195(1-4):49--73.

\bibitem[Mart{\'\i}nez et~al.,
  2017]{Martinez2017NearSurfaceMarsAtmos_VikingToCuriosity}
Mart{\'\i}nez, G., Newman, C., De~Vicente-Retortillo, A., Fischer, E., Renno,
  N., Richardson, M., Fair{\'e}n, A., Genzer, M., Guzewich, S., Haberle, R.,
  et~al. (2017).
\newblock The modern near-surface martian climate: A review of in-situ
  meteorological data from viking to curiosity.
\newblock {\em Space Science Reviews}, 212(1-2):295--338.

\bibitem[Nagaraja et~al., 2020]{Nagaraja2020ExosphereOfMars_MENCA-MOM}
Nagaraja, K., Basuvaraj, P.~K., Chakravarty, S.~C., and Kuttanpillai, P.~K.
  (2020).
\newblock Study of exospheric neutral composition of mars observed from indian
  mars orbiter mission.
\newblock {\em New Astronomy}, 77:101349.

\bibitem[Nier and McElroy, 1976]{Nier1976StrucOfMNUA_NMS_Viking1-2}
Nier, A.~O. and McElroy, M.~B. (1976).
\newblock Structure of the neutral upper atmosphere of mars: Results from
  viking 1 and viking 2.
\newblock {\em Science}, 194(4271):1298--1300.

\bibitem[Nier and McElroy, 1977]{Nier1977CompStrucOfMUA_NMS-Viking1-2}
Nier, A.~O. and McElroy, M.~B. (1977).
\newblock Composition and structure of mars' upper atmosphere: Results from the
  neutral mass spectrometers on viking 1 and 2.
\newblock {\em Journal of Geophysical Research}, 82(28):4341--4349.

\bibitem[Olsen et~al., 2017]{Olsen2017ACE-MIRsuite_ExoMarsTGO}
Olsen, K., Montmessin, F., Fedorova, A., Trokhimovskiy, A., and Korablev, O.
  (2017).
\newblock Trace gas retrievals for the exomars trace gas orbiter atmospheric
  chemistry suite mid-infrared solar occultation spectrometer.

\bibitem[Owen and Biemann,
  1976]{Owen+Biemann1976CompOfMarsAtmosSurface_Argon36Detection}
Owen, T. and Biemann, K. (1976).
\newblock Composition of the atmosphere at the surface of mars: Detection of
  argon-36 and preliminary analysis.
\newblock {\em Science}, 193(4255):801--803.

\bibitem[Owen et~al., 1977]{Owen1977CompOfMarsAtmosSurface}
Owen, T., Biemann, K., Rushneck, D.~R., Biller, J.~E., Howarth, D.~W., and
  Lafleur, A.~L. (1977).
\newblock The composition of the atmosphere at the surface of mars.
\newblock {\em Journal of Geophysical Research (1896-1977)}, 82(28):4635--4639.

\bibitem[Schneider et~al.,
  2015]{Schneider2015AftermathOfCometSidingSpring_IUVS-MAVEN}
Schneider, N.~M., Deighan, J.~I., Stewart, A. I.~F., McClintock, W.~E., Jain,
  S.~K., Chaffin, M.~S., Stiepen, A., Crismani, M., Plane, J. M.~C.,
  Carrillo-Sánchez, J.~D., Evans, J.~S., Stevens, M.~H., Yelle, R.~V., Clarke,
  J.~T., Holsclaw, G.~M., Montmessin, F., and Jakosky, B.~M. (2015).
\newblock Maven iuvs observations of the aftermath of the comet siding spring
  meteor shower on mars.
\newblock {\em Geophysical Research Letters}, 42(12):4755--4761.

\bibitem[Stewart et~al., 1972]{Stewart_Barth1972StrucOfMUA_UVS-Mariner9}
Stewart, A., Barth, C., Hord, C., and Lane, A. (1972).
\newblock Mariner 9 ultraviolet spectrometer experiment: Structure of mars'
  upper atmosphere.
\newblock {\em Icarus}, 17(2):469 -- 474.

\end{thebibliography}
\bibliographystyle{apalike}
\end{document}